\documentclass[aps,physrev,twocolumn,superscriptaddress]{revtex4-2}
\usepackage{graphicx}
\usepackage{dcolumn}
\usepackage{bm}

\begin{document}

\title{Disentangling Coherent and Incoherent Effects in Superconductor Photoemission Spectra via Machine Learning}%

\author{K. H. Bohachov}%
\email{bogachov.kyrylo@gmail.com}%
\affiliation{G.V. Kurdyumov Institute for Metal Physics, N.A.S. of Ukraine}%

\author{A. A. Kordyuk}%
\homepage{https://kau.org.ua/en/kordyuk}%
\affiliation{G.V. Kurdyumov Institute for Metal Physics, N.A.S. of Ukraine}%
\affiliation{Kyiv Academic University, 36 Vernadsky blvd., Kyiv 03142, Ukraine}%
\affiliation{Leibniz Institute for Solid State and Materials Research, IFW Dresden, D-01171, Dresden, Germany}%

\date{\today}

\begin{abstract}
Disentangling coherent and incoherent effects in the photoemission spectra of strongly correlated materials is generally a challenging problem due to the involvement of numerous parameters. In this study, we employ machine learning techniques, specifically Convolutional Neural Networks (CNNs), to address the long-standing issue of the bilayer splitting in superconducting cuprates. We demonstrate the effectiveness of CNN training on modeled spectra and confirm earlier findings that establish the presence of bilayer splitting across the entire doping range. Furthermore, we show that the magnitude of the splitting does not decrease with underdoping, contrary to expectations. This approach not only highlights the potential of machine learning in tackling complex physical problems but also provides a robust framework for advancing the analysis of electronic properties in correlated superconductors.
\end{abstract}


\maketitle

\section{Introduction}

In 1987, Anderson proposed \cite{anderson1987} that the quasi-two-dimensional nature of high-temperature superconductors (HTSC) and the emergence of superconductivity through doping a Mott insulator would lead to novel local physics, distinct from the delocalized electron behavior described by density functional theory (DFT) in the standard local density approximation (LDA). However, early angle-resolved photoemission spectroscopy (ARPES) experiments revealed Fermi surfaces in YBa$_2$Cu$_3$O$_7$ (YBCO) \cite{R_Liu}, Bi$_2$Sr$_2$CaCu$_2$O$_{8+x}$ (BSCCO), and Nd$_{2-x}$Ce$_x$CuO$_4$ that closely resembled those predicted by DFT calculations. These observations indicated a doubly renormalized one-electron conduction band primarily formed by the Cu $3d_{x^2-y^2}$ and O $2p_x$, $2p_y$ orbitals \cite{O_K_Andersen}. Subsequent studies demonstrated that the Fermi surface obeys the Luttinger theorem \cite{luttinger}, suggesting that localization effects are negligible.

However, based on experimental evidence reported on the absence of bilayer splitting in bilayer BSCCO \cite{ding1996}, Andersen \cite{anderson1997} proposed the idea of confinement, suggesting that in the superconducting state, electrons are confined within the CuO layers, leading to perfect two-dimensional (2D) superconductivity as a signature of non-trivial many-body effects.

With the improvement of the ARPES technique in terms of energy and momentum resolution, as well as experimental statistics, bilayer splitting has been observed first in the overdoped BSCCO \cite{feng2001, chuang2001} and later in the wide doping range \cite{kordyuk2002, Borisenko2003, chuang2004}. 

\begin{figure}[b]\centering
	\includegraphics[width=1 \columnwidth]{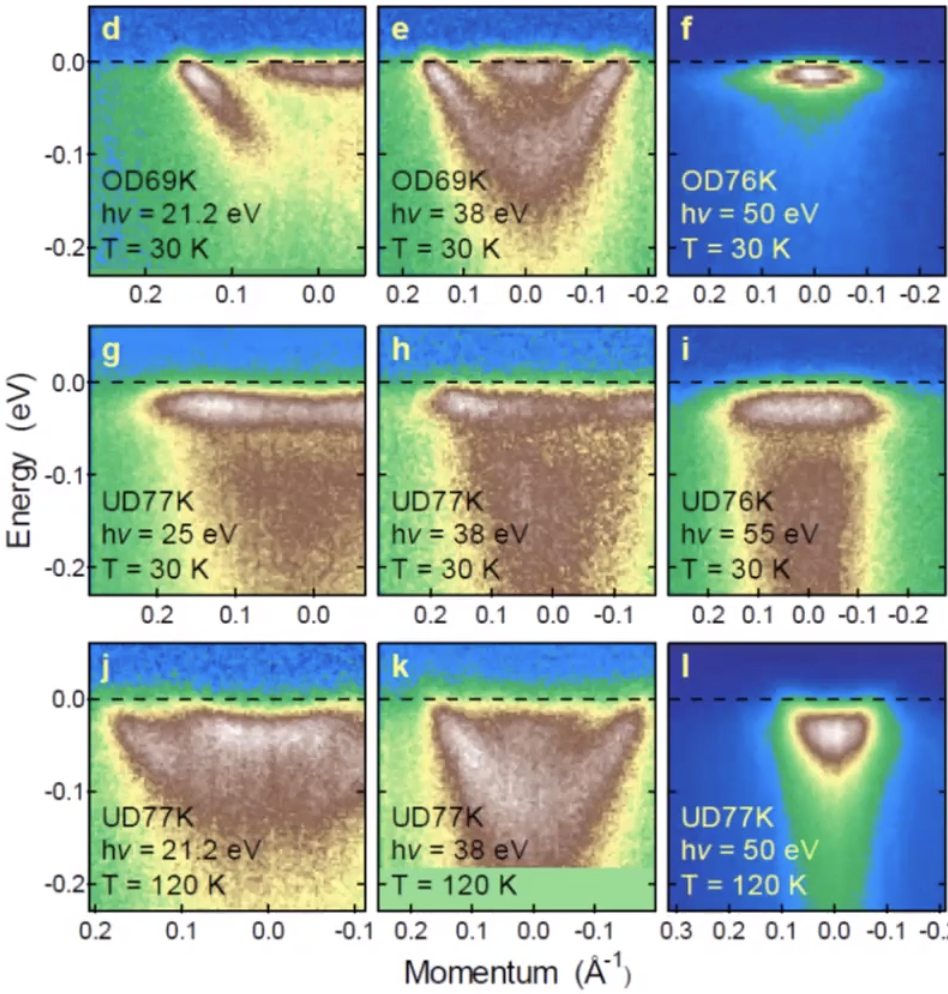}
	\caption{ARPES spectra of bilayer BSCCO along the Brillouin zone boundary, after \cite{Borisenko2003}.}
	\label{arpes_Borisenko}
\end{figure}

Interestingly, despite the evidence mentioned, the splitting in the underdoped region and in the superconducting state remains a point of controversy to this day. The objective reason for this controversy lies in the difficulty of disentangling coherent and incoherent effects in the photoemission spectra of bilayer cuprates---specifically, distinguishing bilayer splitting from the self-energy effect (spectral broadening) caused by the peak in the spin-fluctuation spectrum \cite{Eschrig2006, Dahm2009}. For example, on Figure \ref{arpes_Borisenko}, which shows the ARPES spectra of two-layer BSCCO at the Brillouin zone boundary \cite{Borisenko2003}, one can clearly observe the bilayer splitting at certain excitation energies (here, at 21.2 eV and 38 eV) for the overdoped sample (panels d and e) at any temperature, and for the underdoped sample above $T_c$ (panels i and k). However, it is difficult to see for the underdoped samople below $T_c$ (panels g and h), as well as at the photon energy (here, 50 eV) for which only one of the bands is dominant (panels f, i, l).  

Here we use machine learning methods, specifically convolutional neural networks (CNNs), to address the bilayer splitting issue in BSCCO and its possible evolution with doping. 


\section{Research Methodology\label{submission}}

Neural network training begins with data preparation, which in this case involves photoemission spectra. Achieving good accuracy typically requires a large number of images, which is nearly impossible to obtain from real experiments. Therefore, the idea behind this work was to generate such data using a predefined integral model that combines coherent (band splitting) and incoherent (non-monotonic self-energy) effects caused by the prominent "magnetic resonance" in the spin-fluctuation spectrum \cite{Eschrig2006, Dahm2009}.

\begin{figure}[t]\centering
	\includegraphics[width=0.9\columnwidth]{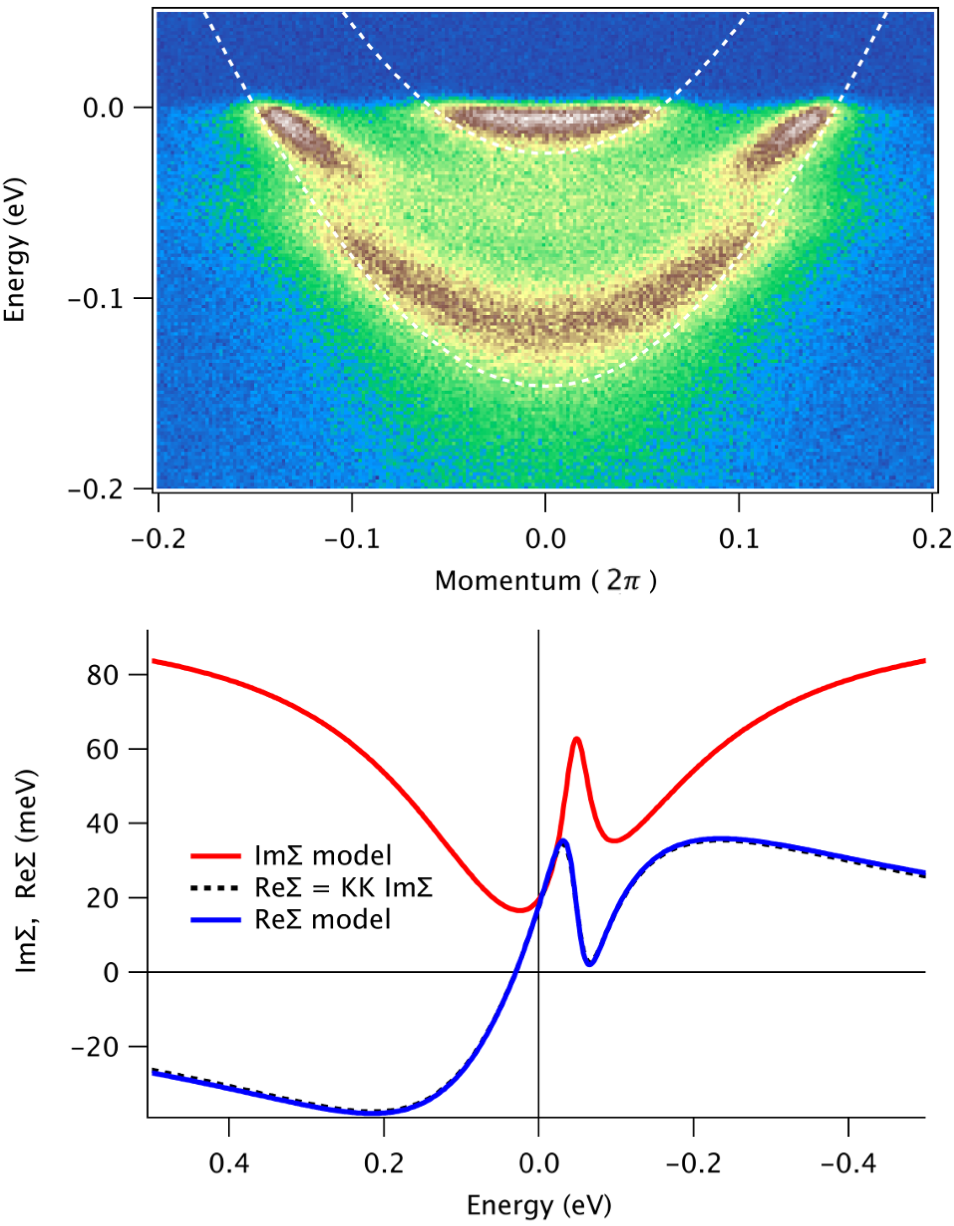}
	\caption{Modeled self-energy parts (bottom): imaginary (red),  real numerical by  Eq.~(\ref{eq6}) (dashed black), and real analytical by Eq.~(\ref{eq8}) (blue); corresponding ARPES spectrum (top), dashed white lines represent the bare dispersions.}
	\label{test_sigma}
\end{figure}

To generate the data, we used a model that assumes the possibility of bilayer splitting into antibonding and bonding bands. In this model, the overall photoemission intensity is expressed as the sum of independent contributions from the two split bands:
\begin{eqnarray}\label{eq1}    I(\omega,k,T)=M_a(h\nu,n,k)f(\omega,T)A_a(w,k)+ \nonumber \\
M_b(h\nu,n,k)f(\omega,T)A_b(w,k),
\end{eqnarray}
where $M_a$ and $M_b$ are the corresponding matrix elements, which mainly depend on the photon energy ($h\nu$), band index ($n$) and momentum ($k$). The function $f(\omega,T)$ represents the Fermi-Dirac distribution, and the single-particle spectral functions are defined similarly for the antibonding and bonding bands:
\begin{equation}\label{eq2}
	A_{a,b}(\omega,k)=\frac{1}{\pi}\frac{\Sigma''(\omega)}{(\omega - \varepsilon_{a,b}(k)-\Sigma'(\omega))^2 +\Sigma''(\omega)^2},
\end{equation}
but with distinct (split) dispersions:
\begin{equation}\label{eq3}
	\varepsilon_{a,b}(k)=\varepsilon^{a,b}_0-t\cos(k),
\end{equation}
while sharing the same self-energy $\Sigma = \Sigma' + i\Sigma''$.

In the first model, we parameterized the imaginary part of the self-energy based on earlier experiments as: 
\begin{equation}\label{eq4}
	\Sigma''(\omega,k)=C+\alpha\omega^2+\beta T^2+\frac{PG^2}{(\omega -\omega_0)^2+G^2},
\end{equation}
while ignoring its real part:
\begin{equation}\label{eq5}
    \Sigma'(\omega,k)=0.
\end{equation}

In the second model, to ensure Kramers-Kronig (KK) consistency,   
\begin{equation}\label{eq6}
\Sigma'(\omega) = \frac{1}{\pi} \text{PV} \int_{-\infty}^{\infty} \frac{\Sigma''(x)}{x - \omega} dx,
\end{equation}
we modified the imaginary part of the self-energy Eq.~(\ref{eq4}) to address the issue of high-energy tails \cite{kordyuk2005}:
\begin{equation}\label{eq7}
\Sigma''(\omega, k) = C + \frac{\alpha \omega^2 \omega^2_c}{\omega_c^2 + \omega^2} + \beta T^2 + \frac{P G^2}{(\omega - \omega_0)^2 + G^2}.
\end{equation}
The corresponding real part can then be expressed analytically as:
\begin{equation}\label{eq8}
\Sigma'(\omega, k) = - \frac{\alpha \omega \omega^3_c}{\omega_c^2 + \omega^2} + \frac{P G (\omega - \omega_0)}{(\omega - \omega_0)^2 + G^2}.
\end{equation}


The analytical dependence of $\Sigma'$ defined by Eq.~(\ref{eq8}) was verified through a numerical calculation with Eq.~(\ref{eq6}) as shown in Figure \ref{test_sigma}.

\subsubsection*{Convolutional Neural Network}

Convolutional neural networks (CNNs) have become the main architecture in computer vision due to their outstanding performance across various visual tasks, including image classification, segmentation, search, and object detection, and have been successfully applied to ARPES data analysis \cite{Ekahana_2023, Imamura_2024, Liu_2023}. 

\begin{figure}[t]\centering
	\includegraphics[width=1\columnwidth]{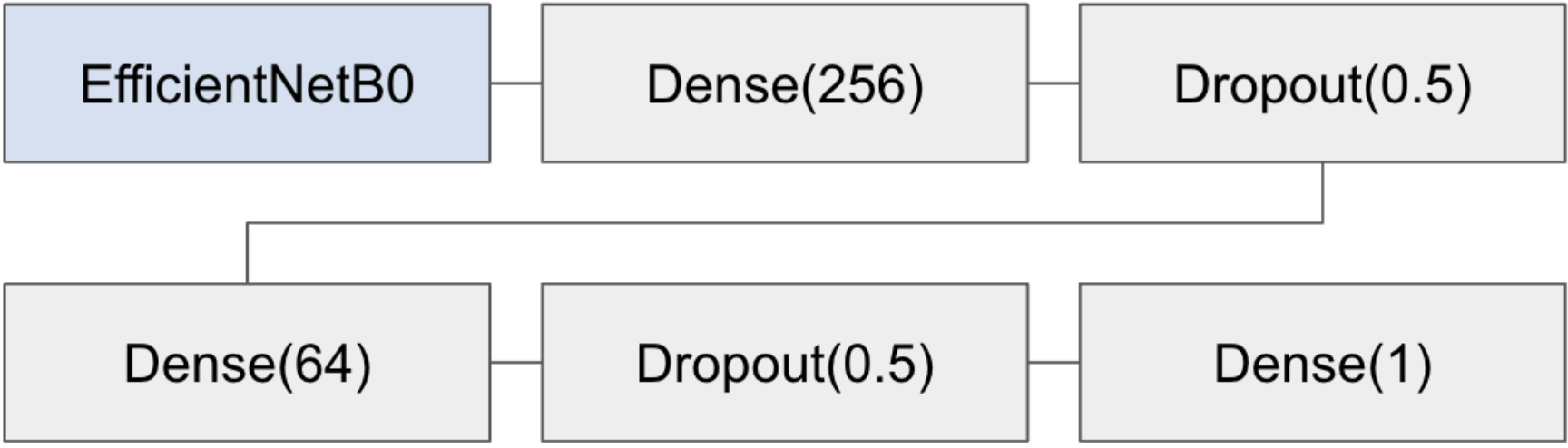}
	\caption{Model architecture}
	\label{pic22}
\end{figure}

The model architecture we use is shown in Figure \ref{pic22}. The main information extractor from the image was the pre-trained on Imagenet convolutional network EfficientNetB0. EfficientNet represents a significant advancement in neural network architecture design by achieving high accuracy with lower computational costs, comparing to previous generation models like ResNet \cite{resnet}. The key innovation of EfficientNet is the use of a compound scaling method that uniformly scales network depth (number of layers), width (number of channels), and resolution (input image size) using a set of fixed scaling coefficients \cite{efficientNet}. The input image size was changed to $224\times224$, and the images were normalized across 3 RGB channels $[0,255]\to[0,1]$. Then hyperparameter search was conducted to find optimal values for dense layers, dropout rates and backbone models.

After the average pooling layer, EfficientNetB0 returns a matrix of dimensions $[batch size, 1280]$, where $batchsize=64$ is the size of one subset of input data. The next layers $Dense(n)$ transform the input features into outputs that the network can use to make a decision, by learning the optimal weights and biases through training, where $n$ is the number of neurons. $Dropout(p)$ is a layer designed for regularization. At each subsequent training iteration, randomly selected nodes are excluded from the network with probability $p$. This allows for a more balanced distribution of weights between neurons \cite{dropout}.

\begin{figure}[b]\centering
	\includegraphics[width=1\columnwidth]{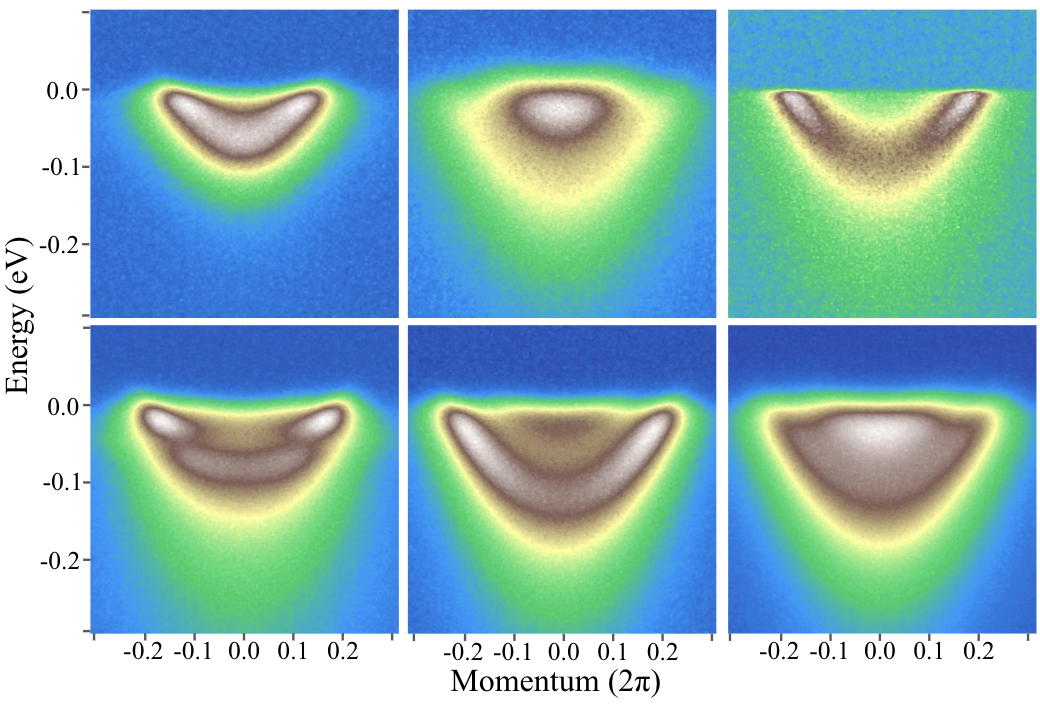}
	\caption{Examples of ARPES spectra generated for training.}
	\label{pic3}
\end{figure}

We employed the mean squared error method (MSE) as the loss function for the primary regression task and cross entropy loss for classification. The role of these loss functions is to compute a quantitative measure that the model strives to minimize during training. Additionally, we utilized $R^2$, F1 score and accuracy as quality metrics. The Adam method with parameters $(\beta_1=0.9, \beta_2=0.999, \alpha=0.0001)$ was used for this minimization \cite{adam}.
At the output, we get a parameter $d\varepsilon_0 = \varepsilon^{a}_0 - \varepsilon^{b}_0$, so exact value of energy splitting for regression task. For classification purposes, we assign labels 0 and 1 to denote spectra with one zone and two zones, respectively.
Data augmentation was applied to reduce overfitting, as it is equivalent to adding new, slightly modified data to the training set. We applied horizontal flip and random resized crop.

\section{Results\label{results}}

\begin{figure}[t]\centering
	\includegraphics[width=1 \columnwidth]{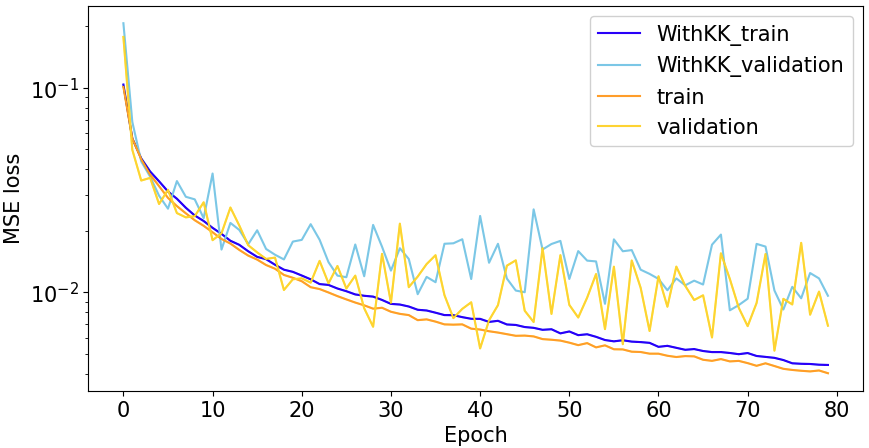}
	\caption{Training and validation MSE loss for experiments with real part of self energy from Kramers-Kronig relations and without during 80 epoch training}
	\label{KK_NoKK_Loss}
\end{figure}

\begin{figure}[b]\centering
	\includegraphics[width=1\columnwidth]{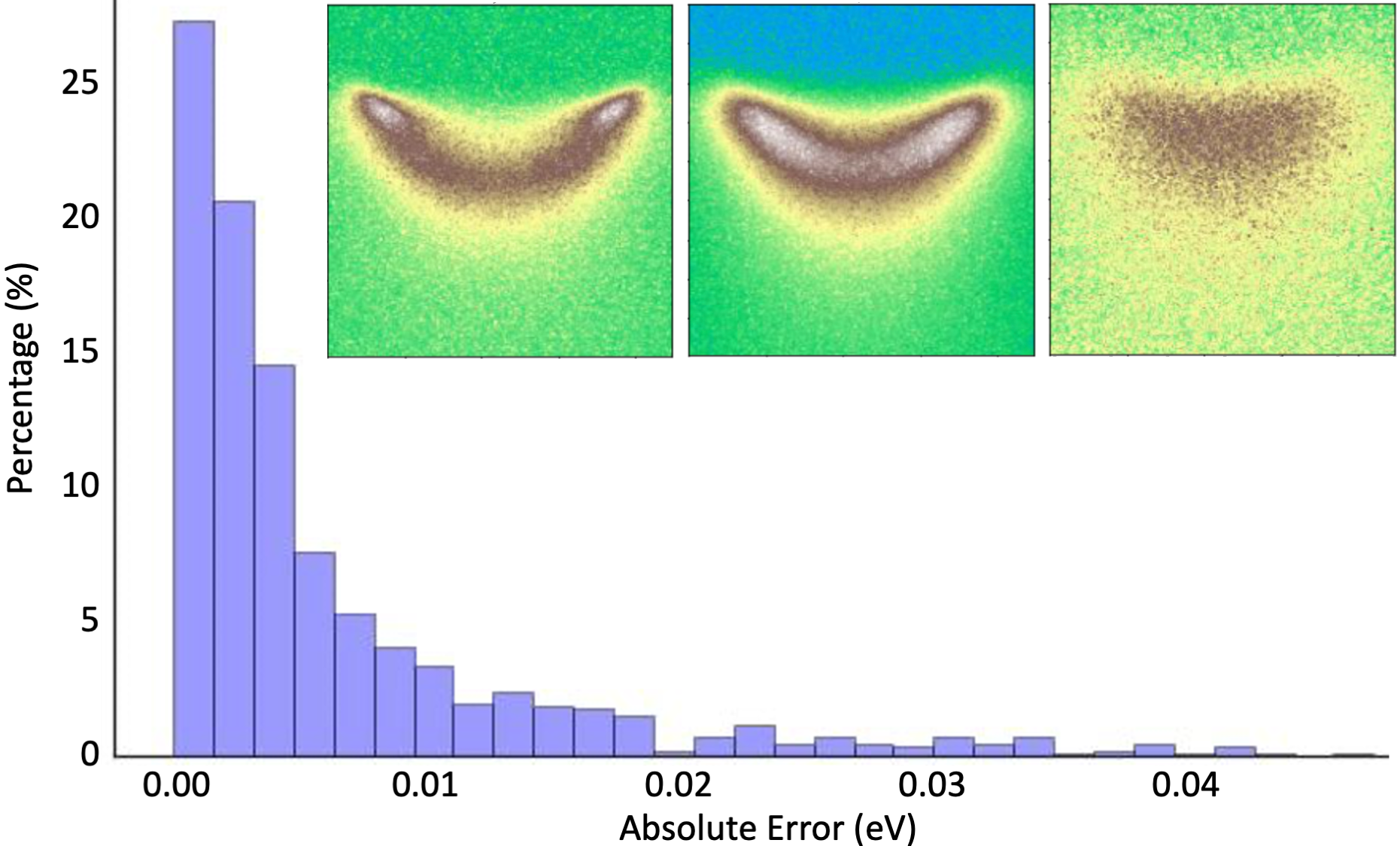}
	\caption{The distribution of absolute errors in percentage for the first model using $d\varepsilon_0 =$ 0.1 eV. The inset shows examples of accurate detection of bilayer splitting using a hierarchical dual-model approach for $d\varepsilon_0$ values of 8, 21, and 89 meV, respectively. The scales are the same as in Fig.~4.}
	\label{hard_case}
\end{figure}

The first experiment focused on classifying the presence of bilayer splitting. Based on the aforementioned model, 10,000 spectra were generated. For half of the data points, either $M_a$ or $M_b$ was set to 0, representing the case without splitting. To evaluate the training quality, the dataset was split into 9,000 images for the training set and 1,000 images for the validation set. Examples of these images are shown in Figure \ref{pic3}.

The training process ran for 50 epochs on a P-100 GPU. This step was crucial point of solving bilayer splitting problem as it helps directly predict presence of the phenomenon in real data. On the validation set, the classification accuracy and F1 score reached $81\%$.

The parameter $d\varepsilon_0$ offers more comprehensive insights into the spectrum structure and the presence of multiple zones. Consequently, we next concentrated on training a regression model to predict this scaled parameter.

\begin{figure}[t]\centering
	\includegraphics[width=1\columnwidth]{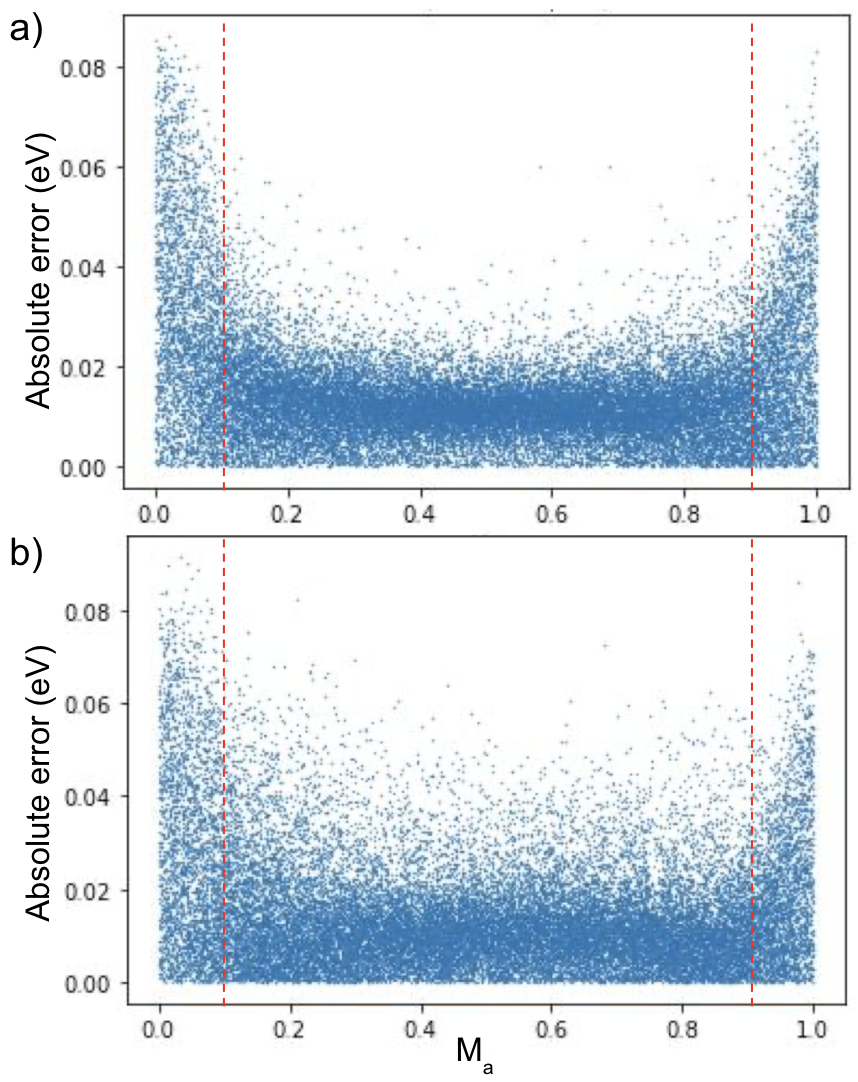}
        \caption{a) Absolute error of $d\varepsilon_0$ splitting in eV for a model trained without the real part of the self-energy, and b) the same but with the inclusion of the real part. Red dashed lines denote the training dataset cutoff.}
	\label{Ma_plot}
\end{figure}

After analyzing numerous spectra with varying levels of Gaussian noise, the noise standard deviation range was narrowed to [0, 50]. This Gaussian noise was synthetically generated and added directly to the spectra to simulate real-world noisy conditions, thereby enhancing the robustness of the model.

Evidently, overly small values of matrix elements can be misleading for generating an effective training dataset. To address this issue, the training data were restricted to the range of $M_a/M_b < 0.1$ and $M_b/M_a < 0.1$, conditions where one zone is at least ten times more intense than the other. This modification improved the final $R^2$ metric by 0.08. However, it was also interesting to examine the model's extrapolation to these limits with much lower intensities.

Returning to model training, we initially employed an approach that excluded the real part of the self-energy. As a result, we generated 24,000 spectra, with 4,800 used for validation. The training, conducted over 80 epochs on a P-100 GPU, achieved a best validation $R^2$ score of 0.87. 

\begin{figure}[t]
	\centering
        \includegraphics[width=0.95\columnwidth]{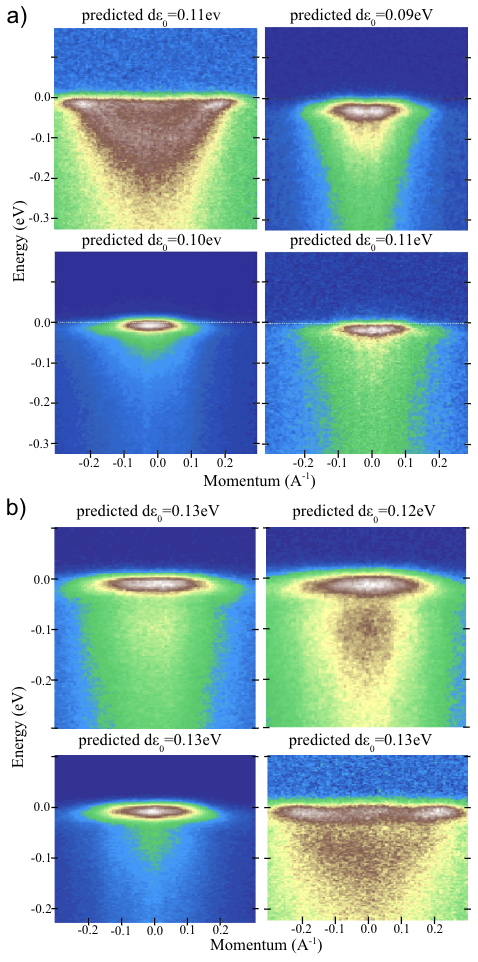}
	\caption{Real data bilayer splitting prediction using a hierarchical dual-model approach: (a) optimal doping, (b) severe underdoping}
	\label{real_results}
\end{figure}

A similar experiment was then conducted using the KK-consistent real part of the self-energy. By replacing Eqs.(\ref{eq4}-\ref{eq5}) with Eqs.(\ref{eq7}-\ref{eq8}), we generated a new dataset to train a neural network with the same architecture and hyperparameters, aiming to test the effect of Kramers-Kronig relations on prediction quality and model fitting. Figure \ref{KK_NoKK_Loss} shows the training and validation curves for both experiments. Overall, no significant improvements in the results were observed.

The distribution of absolute errors in percentage for the first model is presented in Fig.~\ref{hard_case}. The inset shows examples of accurate detection of bilayer splitting for $d\varepsilon_0$ values of 8, 21, and 89 meV, respectively. In this case, we employed a hierarchical approach: initially using binary classification model to verify the presence of both zones, followed by a regression model to predict the specific values of the splitting.

To verify integrity of the model and check results from different angle new dataset was build to perform additional testing. In Fig.~\ref{Ma_plot} we verified symmetry of models behaviour on edge cases of small zone intensities, although models have some bias near 0.1 eV. Best validation $R^2$ reached 0.8.

The presented results illustrate the applicability of an approach in which data for training ML algorithms are generated through modeling based on experimentally determined physical parameters. This method can be applied more broadly to distinguish between possible theoretical models that produce subtle differences in spectra across a range of parameter space. 

Finally, we use the model-trained CNN to determine the bilayer splitting values in real experimental data measured by the IFW Dresden ARPES group at BESSY. We selected spectra for optimally doped ($T_c$ = 90 K, hole concentration $x = 0.16$) and underdoped ($T_c$ = 76 K, $x = 0.10$) bilayer BSCCO measured in the superconducting state at different photon energies, as shown in Fig.~\ref{real_results}. The binary classification model unambiguously predicts the existence of bilayer splitting in all the spectra, and the determined splitting values from the regression model are shown in the figure: 0.10(1) eV for the optimally doped samples and 0.13(1) eV for the underdoped ones. This rules out the possibility that the bilayer splitting vanishes with underdoping.

\section*{Conclusions}

In summary, the detailed analysis of ARPES spectra through our binary classification and regression models demonstrates the effectiveness of CNN training on modeled spectra and confirms earlier findings that establish the presence of bilayer splitting across the entire doping range. We unambiguously show that the magnitude of the splitting does not decrease with underdoping, contrary to some theoretical predictions. Despite the models' high performance metrics, there are still opportunities to enhance sensitivity in low-intensity scenarios and to integrate more accurate physical models. The developed approach not only highlights the potential of machine learning in tackling complex physical problems but also provides a robust framework for advancing the analysis of electronic properties in correlated superconductors.

\begin{acknowledgments}
This work was supported by the German Federal Ministry of Education and Research (BMBF) through the GU-QuMat project (01DK24008) and by the Ministry of Education and Science of Ukraine via grant No. 0124U004211.
\end{acknowledgments}

\bibliography{bohachov}

\end{document}